\documentstyle[prl,version2,aps]{revtex}
\draft
\begin{document}
\begin{title}
LEDUC-RIGHI EFFECT IN SUPERCONDUCTORS
WITH NONTRIVIAL DENSITY OF STATES.
\end{title}

\author{  N.K.Fedorov  }
\begin{instit}
P.N.Lebedev Physical Institute, Russian Academy of Sciences,
Moscow 117924, Russia\\
\end{instit}

\begin{abstract}
 Increasing of electronic thermal conductivity of superconductor
in the model with nontrivial density of states
was considered.
   The electronic thermal conductivity $\kappa (T)$
 of SC in the presence of external magnetic field was
investigated.  It
 was shown that if symmetrical part of
quasiparticle scattering rate less than
 cyclotron energy observed mechanism of increasing  $\kappa (T)$ can be
 separated from  connected with symmetrical electronic
scattering effects another one.
\end{abstract}


{\it Introduction.}
 There are
several ways to provide
the nature of thermal conductivity peak
at temperature below $T_C$
in HTSC
\cite{exp1,exp2}.

Connected with temperature dependence
 of phonon mean free path
increasing of lattice thermal conductivity
 was theoretically
predicted in papers \cite{2}.
Later this approach was applied to
explain the thermal properties of HTSC \cite{3}.
But measurements of
highly  anysotropic thermal current
\cite{4} give evidence that
such method cannot be exhaustive.

Scattering of electrons by nonmagnetic impurities in
anysotropic heavy fermions SC  \cite{5}
and inelastic scattering of
electrons by antiferromagnetic
spin fluctuations \cite{6,7} were taken into account
to understand  the
existence of thermal conductivity peak.

In present paper  the
possible experimental evidence of not connected with
scattering effects mechanisms creating of
thermal conductivity peak is investigated.
As the example of such mechanism can be considered
increasing of electronic thermal conductivity of superconductor
in the model with nontrivial density of states
\cite{my}.

\vspace{4ex}

{\it Thermal conductivity.}
The electronic part of thermal conductivity
can be derived by
 kinetic equation for density matrix
in nonequilibrium technique of
Green-Keldysh functions. It  can be shown
that the charge current
 \begin{equation}
\frac 1e {\bf j}_R(T)=
\langle \sigma _s(p)\rangle _p( \frac{2e}{\hbar c} {\bf
A}_{\bot } - {\bf \nabla} _R\chi (R))-
\langle \eta _p\rangle _p {\bf \nabla}
_RT \eqnum{1} \end{equation}
 and thermal current
 \begin{equation}
{\bf q}_R(T)=\langle \xi _p\sigma _s(p)\rangle _p(
\frac{2e}{ \hbar c} {\bf
A}_{\bot } - {\bf \nabla} _R\chi (R))-
\langle \xi _p\eta _p\rangle _p {\bf
\nabla} _RT \eqnum{2} \end{equation}
 in superconductor, where
 ${\bf A}_{\bot} $ -
vector-potential of electromagnetic field,
 $\chi(R) $ -  order parameters phase.

 Under the condition
\begin{equation}
{\bf j}_R(T)=0
\eqnum{3}
\end{equation}
one can obtain from Eqs.(1,2)
\begin{equation}
{\bf q}_R(T)=   -  \kappa (T)
{\bf \nabla} _RT ,
\eqnum{4}
\end{equation}
where
\begin{equation}
\kappa (T) = \langle \xi _p\eta _p\rangle _p-
\langle \eta _p\rangle
_p\frac{\langle \xi _p\sigma _s(p)\rangle _p}
{\langle \sigma _s(p)\rangle _p}
\eqnum{5} \end{equation}
is the thermal conductivity coefficient.

Eq.(5) without the second term is the Wiedemann-Franz law.
Coefficient
$\langle \xi _p\eta _p\rangle _p$
corresponds to
quasiparticle current from hot edge
of superconductor to cold edge.

To influence to the kinetic properties of SC it is possible
to consider      superconductor with nontrivial
 DOS function:
 \begin{equation}
N_\xi =N_0 + \left\{
\begin{array}{rcl}
N_1,\quad if\quad e_{1}\leq \xi _k\leq e_{2} , \\
N_2,\quad if\quad e_3\leq \xi _k\leq e_4 ;
\end{array}
\right.
\eqnum{6}
\end{equation}
 which is consists of narrow peak $N_1$ not far from $e_F$
with center on $e_0$ and
the wide one $N_2$. The assumption $e_i - \mu(T_C) < 0$ will be
taken into account.
 $d-wave $ symmetry of order parameter is suggested by the
$\Delta(k) = \Delta
\cos(2 \phi)$
, where
$\phi $ - angle from $X-axes$ in Brilluoin zone.
Parameters of the model can be chosen to provide the
 condition  $\mu (0)-\mu (T_c) \sim T_c$.
 Then it is possible to obtain \cite{my} the peak of electronic
thermal conductivity
at $T^*  \sim \left| e_0 - \mu
\right| < T_c$. At this temperature
given the main contribution to the thermal conductivity
electron-hole excitations have energy
$ \sim
  e_0 - \mu(T, \Delta)
$.

The back current of condensate is taken into account by  the
second term Eq.(5).
Consideration of this term in linear approximation
could be incorrect
at temperatures near $T_c$ because of the infinity of
   ${\bf v}_s = ( \frac{2e}{\hbar c} {\bf
A}_{\bot } - {\bf \nabla} _R\chi (R))$
  under the conditions Eq.(3).

Lets consider the case of electronic conductivity
$\frac{\partial J_F}{\partial e_F}>0 ,$ where $J_F$
is the integral over the isoenergy surface in $k$-space near
$e_F$.
 If the DOS has
the narrow peak then
moving of  $\mu(T) $ leads to decreasing of
thermoelectric coefficient
$\left\langle \eta  \right\rangle _p  $  at temperature
$T^* \sim  \left | e_0 - \mu(T^*) \right | $
in accordance with increasing of hole excitations with energy
$ \sim T^* $. The same cause can decreases
 condensate carried heat flow
  $\left\langle \xi _p\sigma _S\right\rangle _p$ can
 at the
temperature  $ \sim T^* $ Therefore besides the coefficient
$\left\langle \eta \right\rangle _p
\frac{\left\langle \xi _p\sigma
_s\right\rangle _p}{\left\langle \sigma _S\right\rangle _p} $
is always positive the relative value of
thermal conductivity peak
$\frac{\kappa (T^{*})}{\kappa (T_C)} $
increases with comparison to Wiedemann-Franz low derived one.
Described mechanism leading to the peak of thermal conductivity
could be
 suppressed
by exponential decreasing of quasiparticle quantity in SC with
isotropic or
had not nods order parameter.

In the case of the hole conductivity
$\left( \frac{\partial J_F}{\partial e_F}<0\right) $
the condensate current decreases peak of electronics
thermal conductivity derived by Wiedemann-Franz low for chosen
DOS .

\vspace{4ex}

{\it Leduc-Righi effect.}
The experimental evidence of electronic nature of thermal
 conductivity
peak could be Leduc-Righi effect \cite{11}.
In SC in external magnetic field
 ${\bf B}\perp \nabla _RT$ (mixed state)
an additional thermal
current occurs.
In this case charge current:
 \begin{equation}
\frac 1e {\bf j}_R(T)  =  <\sigma _S>_{\Vert }{\bf v}_S-
 <\eta >_{\Vert }{\bf \nabla} _RT
 +  <\sigma _S>_{\perp
}\left[ {\bf b}\times {\bf v}_S\right]
 -<\eta >_{\perp }\left[ {\bf b}
\times {\bf \nabla}
_RT\right]
\eqnum{7}
\end{equation}
 and thermal current:
\begin{equation}
{\bf q}_R(T)  =  <\xi \sigma _S>_{\Vert }{\bf v}_S-
<\xi \eta >_{\Vert }{\bf \nabla} _RT
  +  <\xi \sigma
_S>_{\perp }\left[ {\bf b}\times {\bf v}_S\right] -
 <\xi \eta >_{\perp }\left[ {\bf b}\times
{\bf \nabla} _RT\right] ,
\eqnum{8}
\end{equation}
where
${\bf b}=\frac {\bf B}{ B }.$

Eq.(7) under condition Eq.(3)
gives formula for condensate velocity:

\begin{equation}
v_s  =
\frac{
\left\{
\frac{<\eta _p>_{p\Vert }}{<\sigma _s(p)>_{p\Vert }}+%
\frac{<\eta _p>_{p\perp }}{<\sigma _s(p)>_{p\Vert }}
\frac{<\sigma
_s(p)>_{p\perp }}{<\sigma _s(p)>_{p\Vert }}
\right\} }{1+\left(
\frac{%
<\eta _p>_{p\perp }}{<\sigma _s(p)>_{p\Vert }}
\right) ^2} \nabla _RT
 +
\frac{
\left\{
\frac{%
<\eta _p>_{p\perp }}{<\sigma _s(p)>_{p\Vert }}
 -
\frac{<\eta _p>_{p\Vert }}{%
<\sigma _s(p)>_{p\Vert }}
\frac{<\sigma _s(p)>_{p\perp }}{<\sigma
_s(p)>_{p\Vert }}
\right\} }{1+\left(
\frac{%
<\eta _p>_{p\perp }}{<\sigma _s(p)>_{p\Vert }}
\right) ^2}
\left[ b\times \nabla _RT\right]
\eqnum{9}
\end{equation}

From Eqs.(8,9) it is easy to derive followings expressions
for thermal
current and thermal conductivity coefficients:

\begin{eqnarray}
& {\bf q} &  =   - \left\{ <\xi \eta >_{\Vert }-
\frac{<\xi \sigma _S>_{\Vert }}{<\sigma
_S>_{\Vert }}<\eta >_{\Vert }\right\}
{\bf \nabla} _RT
  -  \left\{
<\xi \eta >_{\perp }-\frac{<\xi \sigma _S>_{\Vert }}{<\sigma
_S>_{\Vert }}<\eta >_{\perp } \right\}
 \left[ {\bf b}\times {\bf \nabla} _RT\right] \nonumber \\
& + &
\left\{ <\xi \sigma _S>_{\perp}   -
\frac{<\xi \sigma _S>_{\Vert }}{<\sigma _S>_{\Vert }}
<\sigma _S>_{\perp
} \right\} \frac{<\eta >_{\Vert }}
{<\sigma _S>_{\Vert } }
 \times   \left[ {\bf b}\times {\bf \nabla} _RT\right]
  \equiv  -  \kappa_{\Vert }{\bf \nabla} _RT -
\kappa_{\perp }\left[ {\bf b}\times {\bf \nabla} _RT\right] .
\eqnum{10}
\end{eqnarray}

To derive coefficients $\kappa_{\Vert }, \kappa_{\perp }$
the kinetic equation for SC in
external magnetic field ${\bf B}$
can be used:

\begin{eqnarray}
\frac \partial {\partial t}\widehat{\rho }_0 & + &
\frac 12\left\{ \left[
\frac \partial {\partial {\bf p}}\stackrel{\wedge }{e_p}
;\frac \partial {\partial
{\bf R}}\widehat{\rho }_0\right] _{+}-
\left[ \frac \partial {\partial {\bf R}}\stackrel{%
\wedge }{e_p};\frac \partial {\partial {\bf p}}
\widehat{\rho }_0\right]
_{+}\right\}
+i\left[ \stackrel{\wedge }{e_p};
\widehat{\rho }_0\right] _{-}
= \nonumber \\
&  = & i\omega \stackrel{\wedge }{\delta \rho }
-i\left[\stackrel{\wedge }{e_p};
\stackrel{\wedge }{\delta \rho }\right] _{-}
+\stackrel{\wedge }{\sigma }_Z
\frac e{ c}(\left[ \frac
{\bf p}m\times rot{\bf A}_f\right]
\frac \partial {\partial {\bf p}})\stackrel{\wedge }{\delta
\rho } -
\stackrel{\wedge }{\Sigma }_T\stackrel{%
\wedge }{\delta \rho } ,
\eqnum{11}
\end{eqnarray}
 where $\widehat{\rho }_0$ - equilibrium density matrix,
$\stackrel{\wedge }{\delta \rho }$ -
nonequilibrium correction to
$\widehat{\rho }_0,$
$\stackrel{\wedge }{e_p}=\stackrel{\wedge }{e}
(p \pm \frac ec A_f)$ - energy
matrix,
$\stackrel{\wedge }{\Sigma }_T$ - relaxation energy matrix,
${\bf A}_f$ - vector potential of electromagnetic field in SC.
The solution of Eq.(11) gives
kinetic coefficients \cite{my,10}:

\begin{equation}
\langle \eta _p\rangle _{p \Vert}=\int
\frac{d^3 {\bf p}}{3m(2\pi  \hbar )^3}
\frac{\upsilon _\Vert}{\upsilon _\Vert^2+
\upsilon _{\perp }^2}
(\xi _p - \mu)
 \frac{\partial G(\xi
_p - \mu,T,\triangle (T,p))}{\partial T}
\eqnum{12 a}
\end{equation}

\begin{equation}
\langle \xi _p\eta _p\rangle _{p \Vert}=\int
\frac{d^3 {\bf p}}{3m(2\pi  \hbar )^3}
\frac{\upsilon _\Vert}{\upsilon _\Vert^2
+\upsilon _{\perp }^2}
 (\xi _p - \mu)^2
 \frac{\partial G(\xi
_p - \mu,T,\triangle (T,p))}{\partial T}
\eqnum{12 b}
\end{equation}

\begin{equation}
\langle \eta _p\rangle _{p \perp}=\int
\frac{d^3 {\bf p}}{3m(2\pi  \hbar )^3}
 \frac{
\upsilon _{\perp }}{\upsilon _\Vert^2+\upsilon
_{\perp }^2}
 (\xi _p - \mu)
 \frac{\partial G(\xi
_p - \mu,T,\triangle (T,p))}{\partial T}
\eqnum{12 c}
\end{equation}

\begin{equation}
\langle \xi _p\eta _p\rangle _{p \perp}=\int
\frac{d^3 {\bf p}}{3m(2\pi  \hbar )^3}
 \frac{
\upsilon _{\perp }}{\upsilon _\Vert^2+\upsilon
_{\perp }^2}
 (\xi _p - \mu)^2
 \frac{\partial G(\xi
_p - \mu,T,\triangle (T,p))}{\partial T}
\eqnum{12 d}
\end{equation}

where relaxation rate
$\upsilon _p=\upsilon _0+\upsilon _{\perp }.$
$\upsilon _0$
corresponds to the symmetrical scattering of electrons,
$\upsilon _{\perp }=v_F\sigma _V\frac m{\hbar}\omega _B$;
$\sigma _V$
   is effective temperature dependent
transport cross-section, correspondent to asymmetric
scattering from vortices.
$v_F\sigma _V\frac m{\hbar}\rightarrow 1$
$(T\rightarrow T_C).$

\begin{eqnarray}
G(\xi _k - \mu,T) & = & \frac 12-\frac12
\frac{\xi _k - \mu}{\sqrt{(\xi
_k - \mu)^2+\Delta (T)^2}}
 \times  \tanh \left( \frac{\sqrt{(\xi _k - \mu)^2+\Delta
(T)^2}}{2T}\right);
\nonumber  \\
F(\xi _k - \mu,T) & = & \frac 12
\frac{\Delta (T)}{\sqrt{(\xi _k - \mu)^2+\Delta
(T)^2}}
 \times  \tanh \left( \frac{\sqrt{(\xi _k - \mu)^2+
\Delta (T)^2}}{2T}\right)
\nonumber
\end{eqnarray}
 are integrated over frequency $\omega$ Green-Keldysh functions
$G^{-+}, F^{-+}$.

Correspondent superconducting kinetics coefficients:

\begin{eqnarray}
\langle \sigma _s(p)\rangle _{p \Vert} & = &
\frac1{m}
\int
\frac{d^3 {\bf p}}{m(2\pi  \hbar )^3}
 {\bf p}^2
 \frac{2(\xi -\mu )\left( 4(\xi -\mu )^2+
\upsilon _\Vert^2-
\upsilon _{\perp } ^2\right) }{\left( 4(\xi -\mu )^2
+\upsilon
_\Vert^2-\upsilon _{\perp }^2\right) ^2+
4\upsilon _\Vert^2
\upsilon _{\perp }^2} \nonumber \\
& \times &
\frac{\partial F(\xi _p -
\mu,T,\triangle (T,p))}{\partial \xi _p}
\Delta (T,p)
\eqnum{13 a}
\end{eqnarray}

\begin{eqnarray}
\langle \xi _p\sigma _s(p)\rangle _{p \Vert} & = &
\frac1{m}
\int
\frac{d^3 {\bf p}}{m(2\pi  \hbar )^3}
 {\bf p}^2
 \frac{2(\xi -\mu )\left( 4(\xi -\mu )^2+
\upsilon _\Vert^2-
\upsilon _{\perp }^2\right) }{ \left( 4(\xi -\mu )^2+
\upsilon_{\Vert}^2-\upsilon _{\perp }^2 \right)^2
+4 \upsilon _{\Vert}^2
\upsilon _{\perp }^2 }
(\xi _p - \mu)  \nonumber \\
& \times &
\frac{\partial F(\xi _p -
\mu,T,\triangle (T,p))}{\partial \xi _p} \Delta (T,p)
\eqnum{13 b}
\end{eqnarray}

\begin{eqnarray}
\langle \sigma _s(p)\rangle _{p \perp} & = &
\frac1{m}
\int
\frac{d^3 {\bf p}}{m(2\pi  \hbar )^3}
 {\bf p}^2
\frac{4(\xi -\mu )\upsilon _{\Vert}
\upsilon _{\perp }}{%
\left( 4(\xi -\mu )^2+\upsilon _{\Vert}^2-
\upsilon _{\perp }^2\right)
^2+4\upsilon _\Vert^2\upsilon _{\perp }^2}
\nonumber \\
& \times  &
\frac{\partial F(\xi _p -
\mu,T,\triangle (T,p))}{\partial \xi _p}
\Delta (T,p)
\eqnum{13 c}
\end{eqnarray}

\begin{eqnarray}
\langle \xi _p\sigma _s(p)\rangle _{ p \perp} & = &
\frac1{m}
\int
\frac{d^3 {\bf p}}{m(2\pi  \hbar )^3}
 {\bf p}^2
\frac{4(\xi -\mu )\upsilon _{\Vert}
\upsilon _{\perp } }{%
\left( 4(\xi -\mu )^2+\upsilon _\Vert^2-
\upsilon _{\perp }^2\right)
^2+4\upsilon _\Vert^2\upsilon _{\perp }^2}
 (\xi _p - \mu) \nonumber \\
& \times &
\frac{\partial F(\xi _p - \mu,T,
\triangle (T,p))}{\partial \xi _p}\Delta
(T,p)
\eqnum{13 d}
\end{eqnarray}

In paper \cite{11} following equations for relaxation rates
were used
\cite{12}:
$\upsilon _{\Vert }=\frac{v_F}{l_0}+
v_F\sigma _{tr}\frac m{\hbar}\omega _B,$
$\upsilon _{\perp }=v_F\sigma _{V}\frac m{\hbar}\omega _B .$
 Here is suggested that
 $v_F\sim 10^5m/c ,$
 $\sigma _{tr}\sim 10^{-8}m ,$
 $l_0\sim 10^{-7}m ,$
$B \sim 10 T .$
After this it is possible to rewrite in
Leduc-Righi coefficient
$\langle \xi _p\eta _p\rangle _{p \perp} $
(assumption that
$\frac{\sigma _{tr}}{\sigma _V}$
has the T-independent value is taken into account
\cite{11}):

\begin{equation}
\frac{\upsilon _{\perp }}{\upsilon _{\Vert }^2+\upsilon _{\perp }^2}%
\sim \frac 1{\upsilon _{\perp }}-\frac 1{\upsilon _{\perp }}\frac{%
\upsilon _{\Vert }^2}{\upsilon _{\perp }^2}
\stackrel{ (T\rightarrow T_C) }{\longrightarrow }
\frac 1{\omega _{_{_B}}}-\frac 1{\omega _{_{_B}}}\frac{\upsilon
_0^2}{\omega _B^2}
\eqnum{14}
\end{equation}

First term of right side Eq.(14) corresponds to asymmetric
scattering of electrons from vortices (Lorenz force in normal state).
Second
term
 ($\frac{\upsilon _{\Vert }}{\omega _B} \sim
\frac{\upsilon _0}{\omega _B} \sim 10^{-1}$)
is
connected with electronic scattering. Presented
estimates shows that  in pure SC
($\frac{\upsilon _0}{\omega _B} \ll 1$) with
short enough transport cross section
$\sigma _{tr}$  it is possible to neglect  in
$\kappa_{\perp }$ relaxation
rate
$\upsilon _{\Vert }$ (and T-dependent  $l _0$)
 in comparison
to cyclotron energy
 $\omega _B (\upsilon _{\perp} \sim 1 \div 10^{-1}
\times \omega _B).$

\vspace{4ex}

 {\it Conclusions.}
The temperature dependence of
chemicals potential in the model with nontrivial DOS can leads
to the peak of electronic thermal
conductivity of superconductor.

The possible experimental evidence of the observed
mechanism creating
of thermal conductivity peak could be connected with
measurements of
perpendicular to temperature gradient thermal current
in magnetic field.

\vspace{4ex}

{\it Acknowledgments.}
Author acknowledges useful discussions with
P.I.Arseyev and B.A.Volkov.

This research is supported by
Program "Superconductivity"
  Grant  N96-081 .


\end{document}